\pdfoutput=1

\documentclass[10pt]{sig-alternate-05-2015}

\usepackage[
  pass,
]{geometry}

\usepackage{amsmath}
\usepackage{graphicx}
\usepackage{booktabs}
\usepackage{amssymb}
\usepackage{latexsym}	
\usepackage{array}		
\usepackage{multirow}
\usepackage{subfigure}
\usepackage{algorithm}
\usepackage{algpseudocode}
\usepackage{threeparttable}
\usepackage{paralist}   
\usepackage{xspace}
\usepackage{color}
\usepackage{adjustbox}
\usepackage{balance}
\usepackage{wasysym}

\usepackage{flushend}   

\newif\ifACM
\ACMtrue  

\ifACM
\newcommand{\myfig}{Figure\xspace}
\else
\newcommand{\myfig}{Fig.\xspace}
\fi

\newcommand\x{\textcolor[rgb]{1.00,0.00,0.00}{$\times$}\xspace}

\newcommand{\red}[1]{\textcolor[rgb]{1.00,0.00,0.00}{#1}}
\newcommand{\blue}[1]{\textcolor[rgb]{0.00,0.00,1.00}{#1}}

\begin{document}
\setcopyright{acmcopyright}

\doi{http://dx.doi.org/10.1145/2842665.2843560}

\isbn{978-1-4503-4066-3/15/12}

\conferenceinfo{CoNEXT Student Workshop'15,}{December 1, 2015, Heidelberg, Germany.}

\acmPrice{\$15.00}

\title{\name: Monitoring Per-app Network Performance\\ with Zero Measurement Traffic}
\newcommand{\name}{MopEye\xspace}  

\author{
\alignauthor
Daoyuan Wu$^\S$\titlenote{Most work by this author was performed at HK PolyU.}, Weichao Li$^\dag$, Rocky K. C. Chang$^\dag$, and Debin Gao$^\S$\\
       \affaddr{$^\S$School of Information Systems, Singapore Management University}\\
       \affaddr{$^\dag$Department of Computing, The Hong Kong Polytechnic University}\\
       \email{$^\S$\{dywu.2015, dbgao\}@smu.edu.sg, $^\dag$\{csweicli, csrchang\}@comp.polyu.edu.hk}\\
       \affaddr{\small \red{This is the copy of our CoNEXT'15 poster~\cite{MopEyePoster15}, published in \textbf{December 1, 2015}.}\\\blue{MopEye has been on Google Play for months at \texttt{https://play.google.com/store/apps/details?id=com.mopeye}.}}
}

\maketitle

\begin{abstract}

Mobile network performance measurement is important for understanding
mobile user experience, problem diagnosis, and service comparison.  A
number of crowdsourcing measurement apps (e.g.,
MobiPerf~\cite{mobiperf_android,Mobilyzer15} and
Netalyzr~\cite{netalyzr_android,Layer815}) have been embarked for the
last few years.  Unlike existing apps that use active measurement
methods, we employ a novel passive-active approach to continuously
monitor per-app network performance on unrooted smartphones without
injecting additional network traffic. By leveraging the
\texttt{VpnService} API on Android, \name, our measurement app,
intercepts all network traffic and then relays them to their
destinations using socket APIs.
Therefore, not only \name can measure the round-trip time accurately, it can do so without injecting additional traffic.
As a result, the bandwidth cost (and
monetary cost of data usage) for conducting such a measurement is
eliminated, and the measurement can be conducted free of user
intervention.
Our evaluation shows that \name's RTT measurement is very close to result of
\texttt{tcpdump} and is more accurate than MobiPerf.  We
have used \name to conduct a one-week measurement revealing
multiple interesting findings on different apps' performance.

\end{abstract}

%

\section{Introduction}
\label{sec:intro}

\begin{figure}[t!]
\begin{adjustbox}{center}   
  \subfigure[\small An all-app view.] {
	\label{fig:mainUI}
    \includegraphics[width=0.235\textwidth]{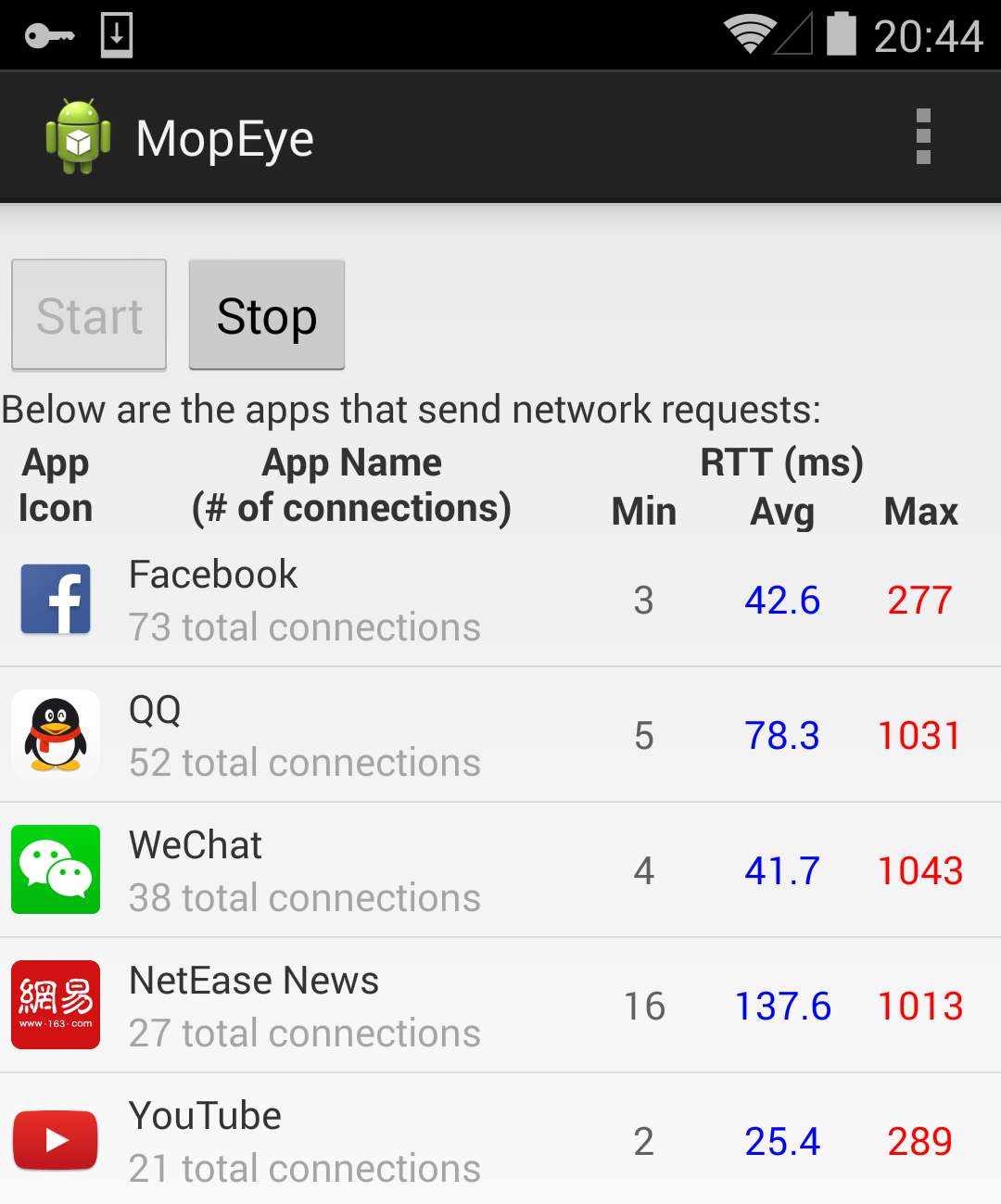}
  }
  \subfigure[\small An individual-app view.] {
	\label{fig:secondUI}
    \includegraphics[width=0.235\textwidth]{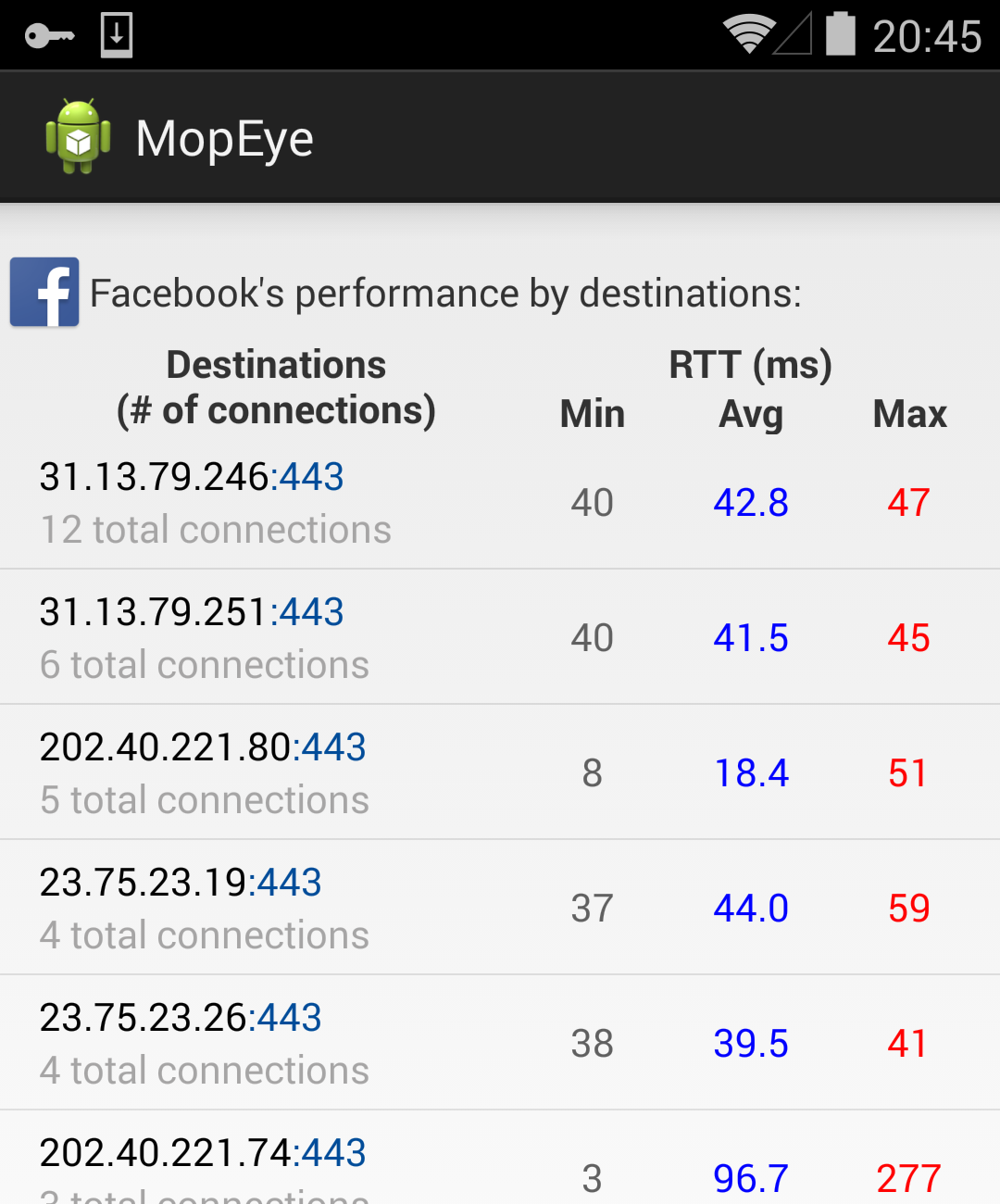}
  }
\end{adjustbox}
\caption{\name's user interfaces.}
\label{fig:toolUI}
\end{figure}

In this paper, we propose a novel passive-active method to
continuously monitor per-app network performance on unrooted
smartphones without injecting additional network
traffic. Specifically, we utilize the \texttt{VpnService} API
available on both Android~\cite{AndroidVPNAPI} and
iOS~\cite{programIOSVPN} platforms for the measurement and implement
it in \name (MObile Performance Eye), our Android measurement
app. With the \texttt{VpnService} service, \name passively captures
the traffic initiated by all apps and the return traffic. \name
then forwards the captured IP packets to the remote servers
using socket calls. Since \name directly communicates with the remote
servers, it can estimate the round-trip time (RTT) based on the socket
calls.  As a result, no additional network traffic is ever incurred.
\name also does not need the root privilege which is required for the
\texttt{tcpdump}-based passive measurement.  \myfig~\ref{fig:toolUI}
shows the two most important user interfaces of \name.

\section{\name Overview}
\label{sec:overview}

\begin{figure}[t!]
\hspace{2ex}
\centering
\includegraphics[width=0.48\textwidth]{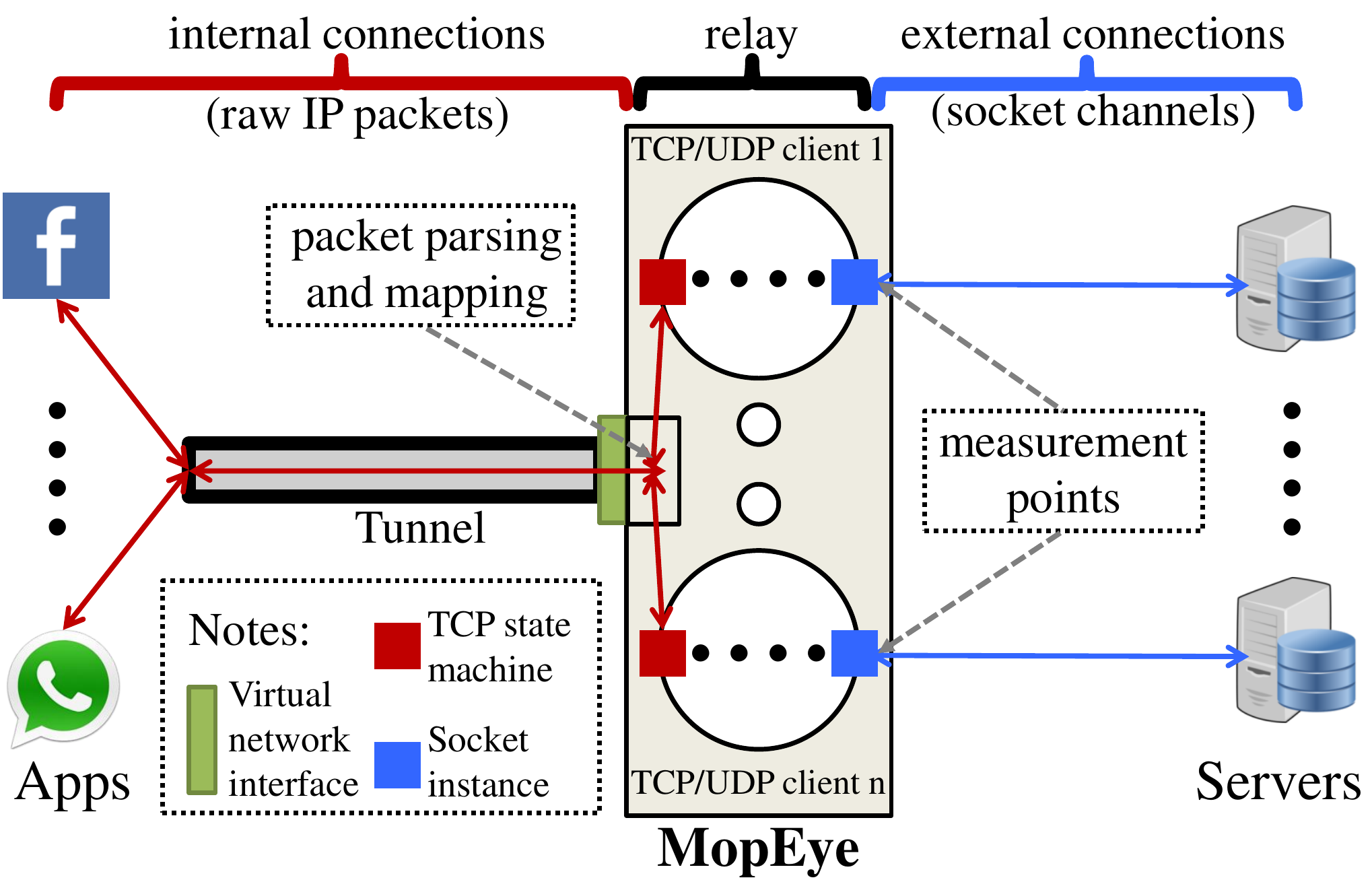}
\caption{An overview of \name.}
\label{fig:overview}
\end{figure}

\myfig~\ref{fig:overview} presents an overview of \name. Using the
Facebook app as an example, we walk through the main steps for \name
to measure its network performance.

\begin{compactenum}

\item \textbf{Packet capturing.} We leverage Android's
  \texttt{VpnService} APIs to build a virtual network interface (green
  box in \myfig~\ref{fig:overview}). The interface enables \name to
  intercept all incoming and outgoing traffic of Facebook.

\item \textbf{Packet parsing and mapping.} \name then parses the
  captured packets to obtain the source and destination IP and TCP/UDP
  headers. Moreover, \name identifies the corresponding app for
  per-app measurement.
    
\item \textbf{Packet relaying.} \name works like a proxy in relaying
  traffic.  As Figure~\ref{sec:overview} shows, \name maintains two
  separate connections with the Facebook server and the app,
  respectively. For the external connection with the server, \name
  allocates a TCP client object in memory and uses its socket instance
  to communicate with the server. For the internal connection with the
  app, as no TCP headers can be retrieved from the socket APIs, \name
  creates its own user-space TCP stack to manage a state machine, and then
  assembles and forwards packets to the tunnel. \name splices the two
  by cross-referencing each TCP client object and its state machine.
    
\item \textbf{Measurement methodology.} \name measures the RTT between
  itself and the Facebook server based on the sockets calls made for
  the external connection. Specifically, \name computes the
  \texttt{SYN-ACK} RTT from the \texttt{connect()} call. Experimental
  results show that only this call can accurately and stably reflect a
  single round of packet exchange.

\item \textbf{Measurement results.} As shown in
  \myfig~\ref{fig:toolUI}, \name displays the RTT results in all-app
  and individual-app views. It shows the number of connections made
  for each app since the beginning of the monitoring and reports the
  minimum, maximum, and mean RTTs.

\end{compactenum}

We have overcome a number of challenges in the design and
implementation of \name, which cannot be elaborated here due to the
space limitation.

\section{Evaluation}
\label{sec:evaluate}

\subsection{Measurement Accuracy and Overhead}
\label{sec:validate}

\textbf{Measurement accuracy.}  We compare \name with MobiPerf~v3.4.0
(the latest version at the time of our evaluation), which is powered
by state-of-the-art Mobilyzer library~\cite{Mobilyzer15}.  We use
MobiPerf's HTTP ping~\cite{wchli15} for comparison because it also
uses \texttt{SYN-ACK} for RTT measurement.  We run \texttt{tcpdump} to
provide the reference measurement results.  In
Table~\ref{tab:VSmobiperf}, we present three sets of results for
Google, Facebook, and Dropbox, which have different ranges of RTTs.
Each result is the mean of ten independent runs because MobiPerf does
not provide detailed results of each run. The difference between
results of \texttt{tcpdump} and \name/MobiPerf is denoted by $\delta$.
This table clearly shows that \name has a much better accuracy than
MobiPerf --- \name's measurement deviates from \texttt{tcpdump}'s by
at most 1ms whereas MobiPerf's deviation ranges from 12ms to 79ms.

By comparing MobiPerf's codes with our \name's implementation, we identify three main contributing factors for \name's higher accuracy. First, \name uses the low-level socket \texttt{connect()} call, instead of the HTTP-level \texttt{HttpURLConnection.connect()} in MobiPerf. Second, the timestamps for \name are collected just before and just after the \texttt{connect()} call. In contrast, MobiPerf's measurement includes the overhead of invoking pre-connect functions, such as \texttt{openConnection()}. Third, \name employs the more accurate nanosecond-level timestamp method, rather than the millisecond-level \texttt{currentTimeMillis()} in MobiPerf.

\begin{table}[t!]
\vspace{-1ex}
\caption{\small The measurement accuracy comparison.}
\hspace{3ex}
\scalebox{
0.85
}{
\begin{adjustbox}{center}
\begin{threeparttable}
\begin{tabular}{ |c | c | c | c | c | c | c | c | c |}

\hline
\multirow{3}{*}{Destinations} & \multicolumn{3}{c|}{\name (mean, in ms)} & \multicolumn{3}{c|}{MobiPerf (mean, in ms)}  \tabularnewline
\cline{2-7}
& \texttt{tcp}   & Mop    & \multirow{2}{*}{$\delta$} &  \texttt{tcp}    & Mobi    &  \multirow{2}{*}{$\delta$}  \tabularnewline
& \texttt{dump}   & Eye*  &   & \texttt{dump}    & Perf    &  \tabularnewline
\hline
\hline

\multirow{2}{*}{Google} & 4.26 &  4    & \textbf{0}      & 4.29  & 16.4  & \textbf{12.11}   \tabularnewline
\multirow{2}{*}{(\texttt{216.58.221.132})} & 4.47 &  5.5  & \textbf{1.03} & 4.35  & 18.5  & \textbf{14.15} \tabularnewline
                      & 5.32 &  5    & \textbf{0}    &  4.85 & 18    & \textbf{13.15}   \tabularnewline
\hline

\multirow{2}{*}{Facebook}& 36.55 & 37   & \textbf{0.45}    & 36.39 &  59.5  & \textbf{23.11}  \tabularnewline
\multirow{2}{*}{(\texttt{31.13.79.251})}  & 36.55 & 37   & \textbf{0.45}  & 36.72 & 55.2  & \textbf{18.48} \tabularnewline
                      & 38.54 & 38.5   & \textbf{0}   & 46.10 & 63.2  & \textbf{17.10} \tabularnewline
\hline

\multirow{2}{*}{Dropbox}& 284.85 & 284.5   & \textbf{0}     & 361.76 &  409.7  & \textbf{47.94}  \tabularnewline
\multirow{2}{*}{(\texttt{108.160.166.126})} & 390.94 & 391  & \textbf{0.06}    & 388.94 & 411.5  & \textbf{22.56} \tabularnewline
                      & 513.78 & 513.5   & \textbf{0}   & 395.87 & 475.2  & \textbf{79.33} \tabularnewline
\hline

\end{tabular}
\begin{tablenotes}
\item [*] \small We round \name's $\mu$s-level results to $ms$-level, e.g., 4.135ms to 4ms.
\end{tablenotes}
\end{threeparttable}
\end{adjustbox}
}
\vspace{-3ex}
\label{tab:VSmobiperf}
\end{table}

\textbf{Network delay overhead.}  It is important for \name
\textit{not} being a bottleneck to other apps.  We thus
measure the additional delay experienced by other apps when \name is
active.
We measure the overhead of the \texttt{SYN-ACK} packets using our measurement tool that invokes \texttt{connect()} to measure the connection time. By subtracting this delay by the MopEye measurement to the same destination, we obtain the delay overhead. For the data packets, we use the popular Ookla Speedtest app~\cite{speedtest_android} to measure the delay with and without MopEye. Their difference is the overhead introduced by MopEye.
Both experiments are repeatedly run on a Nexus~4 running
Android~5.0.  With 95\% confidence interval, the mean delay overhead of a round
of \texttt{SYN-ACK} packets is 4.15$\sim$5.98ms, and that of
data packets is 1.22$\sim$2.18ms.  Considering that the average RTT of
the AT\&T LTE network is about 75.47ms~\cite{RTTresult12}, we find
the additional delay introduced by \name quite acceptable.

%
%
%

\subsection{One-week Measurement Results}
\label{sec:result}

We performed a one-week measurement on a Nexus~5 in Hong Kong on May 2015.
\name relays a total of 5,598 connections out of which
5,410 are successful connections (188 are nonresponsive ones due to,
e.g., unavailable servers).  Among the 5,410 RTTs collected, 4,025
are over WiFi with the remaining 1,385 over LTE 4G network.

\begin{figure}[t!]
\begin{adjustbox}{center}   
  \hspace{-4ex}
  \subfigure[\small Under WiFi.] {
	\label{fig:result1_wifi}
    \includegraphics[width=.26\textwidth, trim=4 5 15 12, clip]{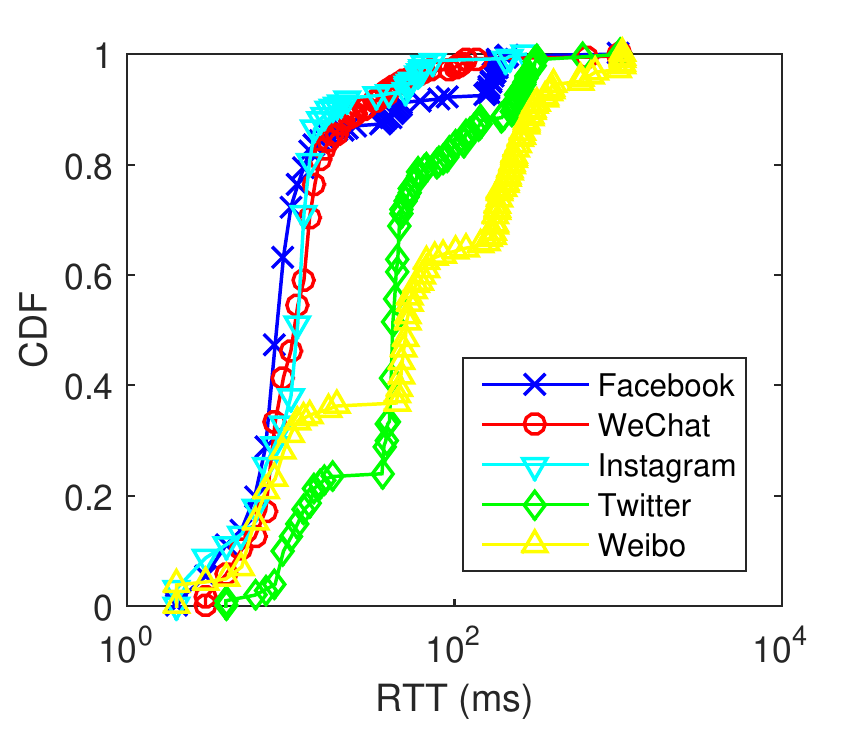}
  }
  \hspace{-3ex}
  \subfigure[\small Under 4G.] {
	\label{fig:result2_4g}
    \includegraphics[width=.26\textwidth, trim=4 5 15 12, clip]{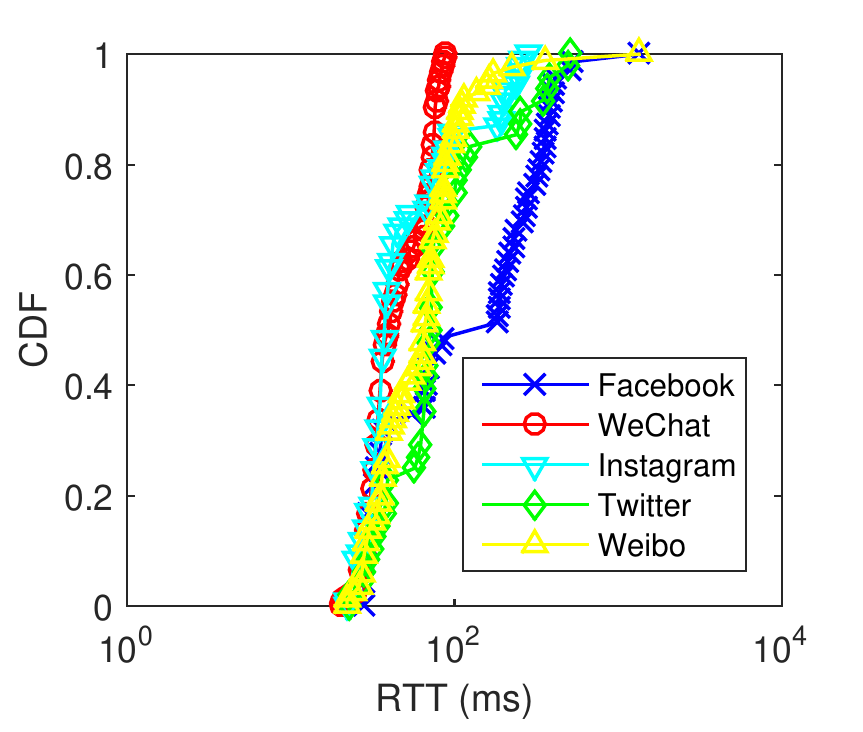}
  }
\end{adjustbox}
\vspace{-4ex}
\caption{\small Top five apps' performance in our dataset.}
\label{fig:result}
\end{figure}

We plot the RTT distribution for the five most popular apps in
Figure~\ref{fig:result1_wifi} (for WiFi) and
Figure~\ref{fig:result2_4g} (for 4G). Facebook, WeChat, and Instagram
achieve better performance than Twitter and Weibo in the WiFi
network, as most of the RTTs ($>$85\%) for these three apps are less
than 20ms.  Over 4G, however, Twitter and Weibo outperform Facebook
which suffers from a significant performance degradation with the
mean RTT risen from 26.1ms to 188.4ms.  To understand why Facebook
experiences such a huge RTT hike in the 4G network, we analyze the
servers connected, and find that there are no local Facebook servers
hosted in Hong Kong for our tested 4G network, whereas local servers
exist for the WiFi network.

As for the 188 nonresponsive connections, they either time out or
fail for other reasons. Most of these connections come from WeChat
and a news app by NetEase (the most popular news app in China).
Our investigation discovers that WeChat's failures are due to their DNS misconfiguration for \texttt{hkminorshort.weixin.qq.com}. When WeChat queries this domain through the smartphone's DNS server (\texttt{8.8.8.8}), the latter therefore responds with \texttt{1.1.1.1} because of its wrong or unregistered DNS configuration. This finding was confirmed and acknowledged by Tencent, and they fixed the problem thereafter. On the other hand, NetEase's failures are due to their extremely large RTTs, causing 18.1\% connections to exceed \name's three-second timeout setting.

\section{Conclusion and Future Work}
\label{sec:conclude}

We proposed a novel measurement app, called \name, to monitor per-app
network performance on unrooted smartphones.
We will deploy \name to Google Play soon for a large-scale measurement study.

\section*{Acknowledgements}

We thank all four anonymous reviewers for their helpful comments.
This work was partially supported by a grant (ref. no. ITS/073/12) from
the Innovation Technology Fund in Hong Kong.

\bibliographystyle{abbrv}
{\small\bibliography{main-2}}  



\end{document}